\documentclass[preprint,showpacs,amsmath,amssymb]{revtex4}

\usepackage{dcolumn}
\usepackage{bm}

\begin{document}


\title{Energy-momentum of a Cosmological Brane Model and the Gauge Hierarchy}

\author{Bei Jia$^{1,2}$, Xi-Guo Lee$^{1,3}$ and Peng-Ming Zhang$^{1,3}$}
\affiliation{\footnotesize{$^1$Institute of Modern Physics,
Chinese Academy of Sciences, P.O.Box 31 Lanzhou, 730000, China\\
$^2$Graduate University of Chinese Academy of Sciences, Beijing, 100080, China\\
$^3$Center of Theoretical Nuclear Physics, National Laboratory of
Heavy Ion Collisions, P.O.Box 31 Lanzhou, 730000, China\\}}


\begin{abstract}
We analyze in this paper the general covariant energy-momentum
tensor of the gravitational system in general five-dimensional
cosmological brane-world models. Then through calculating this
energy-momentum for the cosmological generalization of the
Randall-Sundrum model, which includes the original RS model as the
static limit, we are able to show that the weakness of the
gravitation on the ``visible'' brane is a general feature of this
model. This is the origin of the gauge hierarchy from a
gravitational point of view. Our results are also consistent with
the fact that a gravitational system has vanishing total energy.
\end{abstract}

\maketitle

\section{Introduction}
Since the early attempts of Einstein [1] and others [2], the problem
of the conservation and localization of the energy-momentum of the
gravitational field have been a highly controversial question for
decades. These works share a common disadvantage that their
expressions of the energy-momentum of the gravitational field are
all in the form of pseudotensors, which are not covariant objects
because they inherently depend on the reference frame thus cannot
provide a truly physical local gravitational energy-momentum
density. This difficulty comes from the fact that gravitational
energy-momentum density cannot be locally defined because of the
equivalence principle [3], which states that gravitational field
should not be detectable at a specific point. Duan [4] has proposed
a generally covariant form of the conservation law of
energy-momentum using the orthonormal frames, in which the
energy-momentum is a covariant vector in Riemannian spacetime. It is
generally covariant and is able to overcome the flaws in the
expressions from Einstein and others. There are many advantages of
this form of energy-momentum [5] and we will use it in our
calculations.

The gauge hierarchy problem, i.e. the weakness of the gravitational
interaction compared to other interactions, is one of the most
fundamental problems in theoretical physics nowadays. Randall and
Sundrum proposed a model [6] with one extra dimension which is
compactified as $S^1/\ \mathbb{Z}_2$ with an exponential ``warped''
factor in the metric. They argue that the gauge hierarchy comes from
the fact that the gravitational field is localized at the ``hidden''
or Planck brane and is very weak at our ``visible'' or TeV brane due
to the warped factor. There are numerous works on the generalization
and phenomenological applications of the RS model, such as the
generalization to cosmological issues [7][8]. There is also a
discussion about the energy-momentum of RS model by Liu and others
[9]. Our argument is a generalization of the discussion in [9] and
is mostly based on the cosmological brane-world models.

This paper is arranged as following. In section 2 we first describe
the setup of our work which includes one single extra dimension,
then we briefly review some of the features of the generally
covariant form of the conservation law of energy-momentum proposed
by Duan. After that we use this formulation to calculate the
energy-momentum of our general setup. In section 3 we calculate the
energy-momentum of one particular cosmological brane model which is
a generalization of the original RS model, and discuss the results
in order to show that the weakness of the gravitational field on our
brane is a general feature of this model. Finally, we discuss the
implications of our results.

\section{Energy-momentum from a Five-dimensional Spacetime}
We start with a five-dimensional spacetime with two 3-branes in it.
The fifth dimension which is labeled as $y$ is compactified as
$S^1/\ \mathbb{Z}_2$. The two 3-branes locate at the two fixed
points of the orbifold $y=0$ and $y=\pi$. The total action of this
system is then

\begin{equation} \label{eps}
S=\int d^4x dy \sqrt{-g}\ [2M^3R-\Lambda]+\sum_{i=1,2}\int d^4x
\sqrt{-g^{(i)}}[\mathcal {L}_i-\Lambda_i]
\end{equation}\\
where $g_{MN}$ and $R$ denote the five-dimensional metric and Ricci
scalar respectively, $\Lambda$ and $\Lambda_i$ are the cosmological
constants of that bulk and the branes, and $g^{(i)}_{\mu\nu}$ is the
induced metric on the branes. The uppercase Latin letters $M,N$
stand for the five-dimensional indices. The signature of $g_{MN}$ is
$(-++++)$. We have separated the gravitational part and the matter
part of the action. The general five-dimensional metric from a
cosmological point of view, which means that the 3-branes should be
spatially homogeneous and isotropic, is then

\begin{equation} \label{eps}
ds^2=-n^2(t,y)dt^2+a^2(t,y)\delta_{ij}dx^{i}dx^{j}+b^{2}(t,y)dy^{2}
\end{equation}\\
where we have assumed for simplicity that the usual
three-dimensional space are also spatially flat. The Einstein
tensors are [7]

\begin{equation} \label{eps}
G_{00}=3\biggl[\frac{\dot{a}}{a}\biggl(\frac{\dot{a}}{a}+\frac{\dot{b}}{b}\biggr)
-\frac{n^2}{b^2}\biggl[\frac{a''}{a}+\frac{a'}{a}\biggl(\frac{a'}{a}-\frac{b'}{b}\biggr)\biggr]\biggr]
\end{equation}\\
\begin{equation}
\begin{split}
\label{eps} G_{ii}=&
\frac{a^2}{b^2}\biggl[\frac{a'}{a}\biggl(\frac{a'}{a}+2\frac{n'}{n}\biggr)
-\frac{b'}{b}\biggl(\frac{n'}{n}+2\frac{a'}{a}\biggr)+2\frac{a''}{a}+\frac{n''}{n}\biggr]\\
&
+\frac{a^2}{n^2}\biggl[\frac{\dot{a}}{a}\biggl(-\frac{\dot{a}}{a}+2\frac{\dot{n}}{n}\biggr)
-2\frac{\ddot{a}}{a}+\frac{\dot{b}}{b}\biggl(-2\frac{\dot{a}}{a}+\frac{\dot{n}}{n}\biggr)-\frac{\ddot{b}}{b}\biggr]\\
\end{split}
\end{equation}\\
\begin{equation} \label{eps}
G_{04}=3\biggl(\frac{n'}{n} \frac{\dot{a}}{a}+\frac{a'}{a}
\frac{\dot{b}}{b}-\frac{\dot{a}'}{a}\biggr)
\end{equation}\\
\begin{equation} \label{eps}
G_{44}=3\biggl[\frac{a'}{a}\biggl(\frac{a'}{a}+\frac{n'}{n}\biggr)
-\frac{b^2}{n^2}\biggl[\frac{\dot{a}}{a}\biggl(\frac{\dot{a}}{a}-\frac{\dot{n}}{n}\biggr)+\frac{\ddot{a}}{a}\biggr]\biggr]
\end{equation}\\

Now we turn our attention to the formulation of the energy-momentum
tensor. Here we will use the form proposed by Duan and others [4,5].
According to their argument, the total energy-momentum tensor of the
gravitational system can be derived from the lagrangian of the
gravity in a gravitational theory described by the orthonormal
frames, i.e. the vielbeins. Let's think about a field theory in $n$
dimensional spacetime with the action

\begin{equation} \label{eps}
I=\int d^nx \mathcal{L}(\phi^A,\partial_\mu\phi^A)
\end{equation}\\
where $\phi^A$ represents general fields. If the action is invariant
under a infinitesimal transformations

\begin{align} \label{eps}
x\rightarrow & x^{\mu'} =x^\mu+\delta x^{\mu}\\
\phi^A(x)\rightarrow & \phi'^{A}(x') =\phi^A(x)+\delta \phi^A(x)
\end{align}\\
where we use the Greek letters like $\mu$ to denote the spacetime
coordinates for clarity. If we assume that $\delta \phi^A(x)$
vanishes on the boundary of the spacetime manifold, the it can be
proven that the following result is true [4,5]

\begin{equation} \label{eps}
\partial_\mu(\mathcal{L}\delta x^{\mu}
+\frac{\partial\mathcal{L}}{\partial\partial_\mu \phi^A }\delta_0
\phi^A)+[\mathcal{L}]_{\phi^A}\delta_0 \phi^A=0
\end{equation}\\
where $\delta_0 \phi^A$ is the Lie derivative of $\phi^A$

\begin{equation} \label{eps}
\delta_0
\phi^A=\phi'^{A}(x)-\phi^{A}(x)=\delta\phi^{A}(x)-\partial_\mu\phi^A\delta
x^\mu
\end{equation}\\
while $[\mathcal{L}]_{\phi^A}$ is

\begin{equation} \label{eps}
[\mathcal{L}]_{\phi^A}=\frac{\partial
\mathcal{L}}{\partial\phi^A}-\partial_\mu
\frac{\partial\mathcal{L}}{\partial\partial_\mu \phi^A }
\end{equation}\\
If $\mathcal{L}$ is the total Lagrangian of the system, the we have
$[\mathcal{L}]_{\phi^A}=0$ as the field equation of the field
$\phi^A$. Then from Equation (10) we can get a conservation law
corresponding to infinitesimal transformations (8) and (9)

\begin{equation} \label{eps}
\partial_\mu(\mathcal{L}\delta x^{\mu}
+\frac{\partial\mathcal{L}}{\partial\partial_\mu \phi^A }\delta_0
\phi^A)=0
\end{equation}\\
However, if $\mathcal{L}$ is not the total Lagrangian of the system,
although Equation (10) will still hold, Equation (12) will not.

If we use the vielbein field as the fundamental field in
gravitational theory, the we can decompose $\phi^A$ as
$\phi^A=(e^\mu _a, \psi^B)$, in which $\psi^B$ is an arbitrary
tensor under general displacement transformations and $e^\mu_a$
denotes the vielbeins. $\psi^B$ can always be transformed to a
scalar using vielbeins [4,5], so we can eliminate them from the
gravitational Lagrangian $\mathcal{L}_g$, so that Equation (10)
becomes

\begin{equation} \label{eps}
\partial_\mu(\mathcal{L}\delta x^{\mu}
+\frac{\partial\mathcal{L}_g}{\partial\partial_\mu e^\nu_a }\delta_0
e^\nu_a)+[\mathcal{L}]_{e^\nu_a}\delta_0 e^\nu_a=0
\end{equation}\\
while from the infinitesimal transformations (8) and (9) the Lie
derivative becomes

\begin{equation} \label{eps}
\delta_0e^\mu_a=e^\mu_a\delta(\partial_\nu x^\mu)-(\partial_\nu
e^\mu_a)\delta x^\nu
\end{equation}\\
so we can get

\begin{equation} \label{eps}
\partial_\mu[(\mathcal{L}_g\delta^{\mu}_\sigma
-\frac{\partial\mathcal{L}_g}{\partial\partial_\mu
e^\nu_a}\partial_\sigma
e^\nu_a)+[\mathcal{L}_g]_{e^\sigma_a}e^\mu_a)\delta
x^\sigma+\frac{\partial\mathcal{L}_g}{\partial\partial_\mu
e^\nu_a}e^\sigma_a\delta(\partial_\sigma x^\nu)]=0
\end{equation}\\
which is a general conservation law. If we define

\begin{align} \label{eps}
I^\mu_\sigma=\mathcal{L}_g\delta^{\mu}_\sigma - &
\frac{\partial\mathcal{L}_g}{\partial\partial_\mu
e^\nu_a}\partial_\sigma
e^\nu_a+[\mathcal{L}_g]_{e^\sigma_a}e^\mu_a\\
V^{\mu\sigma}_\nu &
=\frac{\partial\mathcal{L}_g}{\partial\partial_\mu
e^\nu_a}e^\sigma_a
\end{align}\\
then the conservation law (16) becomes

\begin{equation} \label{eps}
\partial_\mu[I^\mu_\sigma \delta x^\sigma+V^{\mu\sigma}_\nu\delta(\partial_\sigma
x^\nu)]=0
\end{equation}\\

The conservation of energy-momentum in special relativity is from
the invariance of the action under the infinitesimal transformation
of the Lorentz coordinates. It can be generated to the general
relativity case as being the general displacement transformation
[4,5]

\begin{equation} \label{eps}
x'^\mu=x^\mu+\delta x^\mu,\   \ \delta x^\mu=e^\mu_ab^a
\end{equation}\\
then the general conservation law above means

\begin{equation} \label{eps}
\partial_\mu(I^\mu_\sigma e^\sigma_a+V^{\mu\nu}_\sigma\partial_\nu
e^\sigma_a)=0
\end{equation}\\
From Einstein equation $\sqrt{-g}T^\mu_a=[\mathcal{L}_g]_{e^a_\mu}$,
where $T^\mu_a$ is the energy-momentum of matter fields, we have
this relation

\begin{equation} \label{eps}
I^\mu_\nu
e^\nu_a=(\mathcal{L}_g\delta^\mu_\nu-\frac{\partial\mathcal{L}_g}{\partial\partial_\mu
e^a_\lambda})e^\nu_a+\sqrt{-g}T^\mu_a
\end{equation}\\
We now define the gravitational energy-momentum as $t^\mu_a$

\begin{equation} \label{eps}
\sqrt{-g}t^\mu_a=(\mathcal{L}_g\delta^\mu_\nu-\frac{\partial\mathcal{L}_g}{\partial\partial_\mu
e^a_\lambda})e^\nu_a+\frac{\partial\mathcal{L}_g}{\partial\partial_\mu
e^\nu_b}e^\sigma_b\partial_\sigma e^\nu_a
\end{equation}\\

Now let's consider the total energy-momentum tensor, which can be
decomposed as [4,5]

\begin{equation} \label{eps}
\Theta^M_a=T^M_a+t^M_a
\end{equation}\\
where $T^M_a$ denotes the energy-momentum tensor of matter while
$t^M_a$ denotes the energy-momentum tensor of the gravitational
field. From now on the lowercase Latin letter $a$ is the index of
the non-coordinate bases from the orthonormal frame. Notice that in
this formulation the energy-momentum tensors are vector-valued
1-forms on a Riemannian manifold, thus the conservation of
energy-momentum can be written in a covariant form from the general
conservation law (21)

\begin{equation} \label{eps}
\nabla_M\Theta^M_a=\frac{1}{\sqrt{-g}}\partial_M(\sqrt{-g}\Theta^M_a)=0
\end{equation}\\
It is shown that there exists a superpotential

\begin{equation} \label{eps}
V^{MN}_a=\frac{\partial \mathcal {L}_g}{\partial(\partial_M\
e^P_b)}\ e^N_be^P_a,\                     V^{MN}_a=-V^{NM}_a
\end{equation}\\
where $e^M_b$ denotes the vielbeins, and $\mathcal {L}_g$ is the
lagrangian of the gravity. We can prove that the total
energy-momentum tensor can be expressed by the superpotential

\begin{equation} \label{eps}
\sqrt{-g}\ \Theta^M_a=\partial_N V^{MN}_a
\end{equation}\\
so that the conservation of energy-momentum can be expressed as

\begin{equation} \label{eps}
\partial_M(\partial_N V^{MN}_a)=0
\end{equation}\\
Then we can calculate the usual energy-momentum four-vector through

\begin{equation} \label{eps}
P_a=\int_{\Sigma}\sqrt{-g}\ \Theta^t_a d\Sigma=\int_S V^{tN}_a dS_N
\end{equation}\\
where the integration hypersurface $\Sigma$ is defined by
$t=constant$, and the second part of the equation is derived from
Gauss's law. Also we can calculate the energy-momentum density

\begin{equation} \label{eps}
\varepsilon_a=\sqrt{-g}\ \Theta^t_a=\partial_N V^{tN}_a
\end{equation}\\

Now we can use this formulation to calculate the energy-momentum of
our brane-world model. The lagrangian of gravity here is

\begin{equation} \label{eps}
\mathcal
{L}_g=\sqrt{-g}[2M^3R-\Lambda]-\sqrt{-g^{1}}\Lambda_1\delta(y-\pi)
-\sqrt{-g^{2}}\Lambda_2\delta(y)
\end{equation}\\
Using the orthonormal frames, this can be expressed as

\begin{equation}\label{eps}
\begin{split}
\mathcal{L}_g=&
\sqrt{-g}\biggl[2M^3(\omega_a\omega^a-\omega_{abc}\omega^{abc})
+\frac{2}{\sqrt{-g}}\partial_M\ (e^M_a\omega^a)-\Lambda\biggr]\\
& -\sqrt{-g^{1}}\Lambda_1\delta(y-\pi)
-\sqrt{-g^{2}}\Lambda_2\delta(y)\\
\end{split}
\end{equation}\\
where $\omega_{abc}$ is the Ricci rotation coefficient defined by
the spin connection $\omega_{Mb}^{a}$ as
$\omega_{abc}=e^M_a\eta_{db}\omega_{Mc}^{d}$ and has the property
$\omega_{abc}=-\omega_{acb}$, while
$\omega_a=\eta^{bc}\omega_{bac}$. It can be proven that the
divergence term in the above equation can be ignored in the
construction of the energy-momentum tensor [4]. The lagrangian which
we really use is then

\begin{equation}\label{eps}
\mathcal{L}_g=2M^3\sqrt{-g}(\omega_a\omega^a-\omega_{abc}\omega^{abc})
-\sqrt{-g^{1}}\Lambda_1\delta(y-\pi)
-\sqrt{-g^{2}}\Lambda_2\delta(y)
\end{equation}\\
Using Equation (26), we can get the expression of the superpotential

\begin{equation}\label{eps}
V^{MN}_a=4M^3\sqrt{-g}\biggl[e^M_be^N_c\omega_a^{bc}+\biggl(e^M_ae^N_b-e^N_ae^M_b\biggr)\omega^b\biggr]
\end{equation}\\
Rewrite the metric in Equation (2) using the orthonormal frame

\begin{equation}\label{eps}
ds^2=-\hat{\theta}^0\otimes\hat{\theta}^0
+\hat{\theta}^1\otimes\hat{\theta}^1
+\hat{\theta}^2\otimes\hat{\theta}^2
+\hat{\theta}^3\otimes\hat{\theta}^3
+\hat{\theta}^4\otimes\hat{\theta}^4
\end{equation}\\
we are able to get the basis of the orthonormal frame

\begin{equation}\label{eps}
\hat{\theta}^a=(n(t,y)dt,a(t,y)dx^1,a(t,y)dx^2,a(t,y)dx^3,b(t,y)dy)
\end{equation}\\
and their components

\begin{equation}\label{eps}
e^0_t=n(t,y),\             e^{i}_{x^i}=a(t,y),\ e^4_y=b(t,y)
\end{equation}\\
Then from the torsion-free condition

\begin{equation}\label{eps}
d\hat{\theta}^a+\omega^a_b\wedge\hat{\theta}^b=0
\end{equation}\\
we are able to find the non-vanishing components of
$\omega_{a}^{bc}$ and $\omega_{a}$

\begin{align}\label{eps}
\omega^{04}_0 & =-\omega^{40}_0=\frac{1}{b}\frac{n'}{n}&
\omega^{04}_4 &=-\omega^{40}_4=\frac{1}{n}\frac{\dot{b}}{b}\nonumber\\
\omega^{i0}_i &=-\omega^{0i}_i=-\frac{1}{n}\frac{\dot{a}}{a}&
\omega^{i4}_i &=-\omega^{4i}_i=\frac{1}{b}\frac{a'}{a}\\
\omega_0 &
=-\frac{1}{n}\biggl(\frac{\dot{b}}{b}-3\frac{\dot{a}}{a}\biggr)&
\omega_4
&=-\frac{1}{b}\biggl(\frac{n'}{n}+3\frac{a'}{a}\biggr)\nonumber
\end{align}\\
The superpotential can be derived from Equation (34)

\begin{align}\label{eps}
V^{ty}_0 & =-V^{y\ t}_0=\frac{-12M^3a^2a'}{b}\nonumber\\
V^{x^it}_i & =-V^{tx^i}_i=\frac{4M^3a^2b}{n}\biggl(\frac{\dot{b}}{b}-4\frac{\dot{a}}{a}\biggr)\\
V^{x^iy}_i &
=-V^{yx^i}_i=\frac{4M^3a^2n}{b}\biggl(-\frac{n'}{n}-2\frac{a'}{a}\biggr)\nonumber
\end{align}\\
Now we can calculate the energy-momentum density from Equation (30).
We find that the only non-zero component of the energy-momentum
density is the energy density

\begin{equation}\label{eps}
\varepsilon_0=-12M^3\biggl(\frac{aa'^2}{b}-\frac{a^2a'b'}{b^2}+\frac{a^2a''}{b}\biggr)
\end{equation}\\
which is obviously independent of n(t,y). Therefore, the only
non-zero component of the energy-momentum four-vector is the total
energy

\begin{equation}\label{eps}
P_0=12M^3\mathcal {V}\biggl(\frac{a^2a'}{b}\biggr)\
\bigg|_{y=\pi}^{y=0}
\end{equation}\\
where $\mathcal {V}$ represents the volume of the three dimensional
space. We will discuss the implications of these results through
analyzing a specific model in the next section.

\section{Energy-momentum of the Cosmological RS Model}
A natural generalization of RS model to the cosmological content has
inflationary property [8], which include the original RS model as
the static limit. Using the notations above, this generalization is
obtained if we set

\begin{equation}\label{eps}
a=A(t)E(y),\qquad n=E(y),\qquad b=B_0
\end{equation}\\
where $B_0$ is a constant which represents the size of the fifth
dimension. Then we can solve the Einstein equations to get the
expressions of $A(t)$ and $E(y)$. Like in RS model, we only consider
the cosmological constants dominated cases, so the Einstein
equations are

\begin{equation} \label{eps}
G_{00}=\frac{3}{E^2}\biggl(\frac{\dot{A}}{A}\biggr)^2
-\frac{3}{B_0^2}\biggl[\biggl(\frac{E'}{E}\biggr)^2+\frac{E''}{E}\biggr]
=\frac{1}{4M^3}\biggl[\Lambda+\frac{\Lambda_1}{B_0}\delta(y)+\frac{\Lambda_2}{B_0}\delta(y-\pi)\biggr]
\end{equation}\\
\begin{equation} \label{eps}
G_{ii}=\frac{1}{E^2}\biggl[\biggl(\frac{\dot{A}}{A}\biggr)^2+2\frac{\ddot{A}}{A}\biggr]
-\frac{3}{B_0^2}\biggl[\biggl(\frac{E'}{E}\biggr)^2+\frac{E''}{E}\biggr]
=\frac{1}{4M^3}\biggl[\Lambda+\frac{\Lambda_1}{B_0}\delta(y)+\frac{\Lambda_2}{B_0}\delta(y-\pi)\biggr]
\end{equation}\\
\begin{equation} \label{eps}
G_{44}=\frac{3}{E^2}\biggl[\biggl(\frac{\dot{A}}{A}\biggr)^2+\frac{\ddot{A}}{A}\biggr]
-\frac{6}{B_0^2}\biggl(\frac{E'}{E}\biggr)^2 =\frac{\Lambda}{4M^3}
\end{equation}\\
where the (04) part of the Einstein tensor automatically vanishes.
The boundary conditions are

\begin{equation} \label{eps}
\frac{E'}{E}\bigg|_{0_-}^{0+}=-\frac{B_0}{12M^3}\Lambda_1\nonumber
\end{equation}\\
\begin{equation} \label{eps}
\frac{E'}{E}\bigg|_{-\pi}^{+\pi}=-\frac{B_0}{12M^3}\Lambda_2
\end{equation}\\

Using Equations (44) and (46), it is shown that there is a
inflationary solution of A(t) [8]

\begin{equation}\label{eps}
A(t)=A_0e^{H_0t},\qquad H_0=\frac{\dot{A}}{A}
\end{equation}\\
Then From Equation (46) we can find the solution to $E(y)$
satisfying the orbifold symmetry [8]

\begin{equation}\label{eps}
E(y)=\frac{H_0}{k}\sinh (-kB_0|y|+c),\qquad
k=\sqrt{\frac{-\Lambda}{24M^3}}
\end{equation}\\
From the boundary conditions (47) we get

\begin{equation} \label{eps}
k_1=k\coth(c),\qquad -k_2=k\coth\biggl(-kB_0\pi+c\biggr)
\end{equation}\\
where $k_i=\Lambda_i/24M^3$. So the metric of this model is

\begin{equation} \label{eps}
ds^2=\biggl(\frac{H_0A_0}{k}\biggr)^2\sinh^2
(-kB_0|y|+c)[-dt^2+e^{2H_0t}\delta_{ij}\ dx^{i}dx^{j}]+B_0^2dy^{2}
\end{equation}\\
Obviously this metric belongs to the ``warped geometry''.

Now we can calculate the energy-momentum density and the
energy-momentum four-vector of this model from Equations (41) and
(42). Again the only non-zero components are the energy density and
total energy at a given time $t$

\begin{equation} \label{eps}
\varepsilon_0=\frac{-12M^3B_0H^3_0A^3_0}{k}e^{-3H_0t}\cosh(-2kB_0|y|+2c)\sinh(-kB_0|y|+c)
\end{equation}\\
\begin{equation} \label{eps}
P_0=\frac{2H^3_0A^3_0M^3}{k^2}e^{-3H_0t}\mathcal
{V}[3\cosh(c)-\cosh(3c)-3\cosh(c-B_0k\pi)+\cosh(3c-3B_0k\pi)]
\end{equation}\\
Let consider the total energy $P_0$ first. It is shown by many
authors that the total energy of a gravitational system is zero
[10], while from Equation (53) we can see that the total energy is
infinity. This might be from the effect of the gravity along the
warped fifth dimension. Notice if $B_0\rightarrow 0$, i. e. the
fifth dimension vanishes, then we have $P_0\rightarrow 0$, which is
consistent with [9].

On the other hand, if we set

\begin{equation}\label{eps}
a=n=e^{-kr|y|},\qquad b=r
\end{equation}\\
then we recover the original RS model as the static limit. In this
case the energy-momentum density and total energy are [9]

\begin{equation} \label{eps}
\varepsilon_0=-36M^3k^2e^{-3kr|y|}
\end{equation}\\
\begin{equation} \label{eps}
P_0=12M^3k(e^{-3kr\pi}-1)\mathcal {V}
\end{equation}\\
If $r\rightarrow 0$ then $P_0\rightarrow 0$, which is consistent
with above discussion. However, Equation (55) indicates that the
energy density, which only includes the energy density of the
gravity along the fifth dimension, is much grater at the Planck
brane than at the TeV brane

\begin{equation} \label{eps}
\frac{\varepsilon_0^{P}}{\varepsilon_0^{T}}=e^{3kr\pi}
\end{equation}\\
This might be a reflection of the gauge hierarchy in RS model that
the gravity is much stronger on the Planck brane than on the TeV
brane, because the exponential difference between the gravitational
energy density on the two branes which means gravity is mostly
localized on the Planck brane [9]. In RS $e^{kr\pi}$ is required to
be of order $10^{15}$ in order to generate the gauge hierarchy, then
we have

\begin{equation} \label{eps}
\frac{\varepsilon_0^{P}}{\varepsilon_0^{T}}\approx 10^{45}
\end{equation}\\
Notice that the energy density here is negative. This might be the
result of the brane model --- the total energy of a gravitational
system is zero, which means if we set the matter energy to be
positive, then the gravitational energy could be negative. In this
brane-world model there is no matter energy to cancel the
gravitational energy along the fifth dimension so that the energy
density here could be pure gravitational energy which is negative.

RS model is only the static limit to our discussion here. Let's
consider now the gauge hierarchy from Equation (52). We can see

\begin{equation} \label{eps}
\frac{\varepsilon_0^{P}}{\varepsilon_0^{T}}
=\frac{\cosh(2c)\sinh(c)}{\cosh(-2kB_0\pi+2c)\sinh(-kB_0\pi+c)}
\end{equation}\\
According to our previous discussion, we can set the energy density
here to be negative. This puts a constrain on the constant $c$ that
it should be larger that $kB_0\pi$ (which we will meet again later
in agreement with the result from [8]), then again the energy
density on the Planck brane could be much greater than on the TeV
brane. Furthermore, if $c$ is near $kB_0\pi$, we can get a extremely
large difference between these two energy densities. Based on our
previous discussion, this again means that we find a hierarchy
between the Planck brane and the TeV brane if
$\varepsilon_0^{P}/\varepsilon_0^{T}\approx 10^{45}$. There is only
one requirement here in order to generate this hierarchy that $c$
should be larger that $kB_0\pi$, which is natural according to our
discussion. The magnitude of the hierarchy is determined by the
difference between $c$ and $kB_0\pi$; whatever the value of the size
of the fifth dimension is, the suitable hierarchy is generated from
choosing the constant $c$. If $B_0$ is determined by some other
mechanism, then we can decide the value of $c$ according to the
gauge hierarchy. Therefore, the gauge hierarchy could be regarded as
a general feature of this generalized RS model, no matter how the
size of the extra dimension is stabilized.

From Equations (50) we can see that the constrain on $c$ is cast
into the constrain on $k_1$ and $k_2$, which is consistent with the
results from [8], which supports our discussion of setting the
energy density to be negative. Our requirement that $c>kB_o\pi$,
which gives a negative energy density, agrees with the second one of
Equations (50) in order to get a negative $k_2$. Indeed, the gauge
hierarchy problem and the cosmological constant problem are recast
into the fine-tuning problem of the bulk and the brane cosmological
constants, whose detail can be found in [8], for instance,
$k\mathrm{csch}(-kB_0\pi+c)=\sqrt{k_2^2-k^2}\lesssim 10^{-60}M_P$.
However, the cosmological constant problem here is dependent upon
the value of $B_0$, i.e. the modulus stabilization, while the gauge
hierarchy may not.

To conclude, we analyze the covariant energy-momentum of a general
cosmological five-dimensional brane model, then specifically of a
generalized RS model which includes the original RS model as the
static limit. The total energy in our case agrees with the general
result from others, which is that the total energy of a
gravitational system is zero, if the extra dimension vanishes. We
are also able to show that the gauge hierarchy is a general feature
of this model from a gravitational point of view, which is
independent from the stabilization of the size of the fifth
dimension. Our results confirms the conclusions from [8] but from a
different point of view.

\begin{acknowledgments}
This work is supported by the CAS Knowledge Innovation Project
(No.KJCX3-SYW- N2,No.KJCX2-SW-N16) and the National Natural Science
Foundation of China (10435080, 10575123, 10604024).
\end{acknowledgments}


\end{document}